\documentclass{emulateapj}

\newcommand{\msun}{{\rm M}_\odot}
\newcommand{\msunyr}{{\rm M}_\odot\ {\rm yr}^{-1}}
\newcommand{\gaea}{\sc{gaea}}
\def\lesssim{\lower.5ex\hbox{$\; \buildrel < \over \sim \;$}}
\def\gtrsim{\lower.5ex\hbox{$\; \buildrel > \over \sim \;$}}

\received{[...]}
\revised{[...]}
\accepted{[...]}

\shorttitle{High-z galaxies in {\gaea}}
\shortauthors{Fontanot, Hirschmann \& De Lucia}

\begin{document}

\title{Strong stellar-driven outflows shape the evolution of galaxies
  at cosmic dawn.}


\author{Fabio Fontanot}
\affil{INAF - Astronomical Observatory of Trieste, via G.B. Tiepolo 11, I-34143 Trieste, Italy}

\author{Michaela Hirschmann}
\affiliation{Sorbonne Universit\'es, UPMC-CNRS, UMR7095, Institut d'Astrophysique de Paris, 75014, Paris, France}

\author{Gabriella De Lucia}
\affiliation{INAF - Astronomical Observatory of Trieste, via G.B. Tiepolo 11, I-34143 Trieste, Italy}

\begin{abstract}
We study galaxy mass assembly and cosmic star formation rate (SFR) at
high-redshift (z$\gtrsim$4), by comparing data from multiwavelength
surveys with predictions from the GAlaxy Evolution and Assembly
({\gaea}) model. {\gaea} implements a stellar feedback scheme
partially based on cosmological hydrodynamical simulations, that
features strong stellar driven outflows and mass-dependent timescale
for the re-accretion of ejected gas. In previous work, we have shown
that this scheme is able to correctly reproduce the evolution of the
galaxy stellar mass function (GSMF) up to $z\sim3$. We contrast model
predictions with both rest-frame Ultra-Violet (UV) and optical
luminosity functions (LF), which are mostly sensible to the SFR and
stellar mass, respectively. We show that {\gaea} is able to reproduce
the shape and redshift evolution of both sets of LFs. We study the
impact of dust on the predicted LFs and we find that the required
level of dust attenuation is in qualitative agreement with recent
estimates based on the UV continuum slope. The consistency between
data and model predictions holds for the redshift evolution of the
physical quantities well beyond the redshift range considered for the
calibration of the original model. In particular, we show that {\gaea}
is able to recover the evolution of the GSMF up to z$\sim$7 and the
cosmic SFR density up to z$\sim$10.
\end{abstract}


\section{Introduction}\label{sec:intro}
Since the introduction of the drop-out technique \citep[see
  e.g.][]{Steidel96}, the study of galaxy populations at increasingly
higher redshift has provided fundamental contributions to our
understanding of the first stages of structure formation in the
Universe. The advent of space based observatories (like the {\it
  Hubble Space Telescope} - HST - and {\it Spitzer}) has allowed us to
push these studies to the $3<z<6$ redshift range and beyond
\citep[e.g.][]{Bouwens16}. Programs like the {\it Great Observatories
  Origins Deep Survey} (GOODS), the {\it Cosmic Assembly Near-infrared
  Deep Extragalactic Legacy Survey} (CANDELS) and the {\it Hubble
  Ultra Deep Field} (HUDF) provide excellent datasets to select high-z
galaxy candidates (see e.g. \citealt{Bouwens15} and
\citealt{Finkelstein15} and references herein). Early work in the
field focus on the determination of the luminosity function (LF) in
the rest-frame Ultra-Violet (UV), easily accessible through optical
photometry (e.g. using the Advanced Camera for Surveys - ACS - onboard
the HST). Since rest-frame UV bands provide information about the
unobscured star formation rate (SFR) for detected sources, the
resulting UV-LF can be used to study the evolution of the cosmic SFR
and to assess the galaxy contribution to the reionization of the
Universe \citep{Robertson13, Fontanot14a}. Later instruments like the
Infrared Array Camera (IRAC) on {\it Spitzer} and the Wide Field
Camera 3 (WFC3) on HST have considerably widened the wavelength range
available for high-z studies. Recently, several groups have used this
multiwavelength information to estimate the LF at rest-frame optical
wavelengths \citep{Stefanon16} and/or the galaxy stellar mass function
(GSMF) at $z>4$ \citep{Gonzalez11, Grazian15}. The rest-frame
optical/near-infrared information complements the rest-frame UV data
so that, when considered together, they represent a powerful tool to
constrain the physical mechanisms shaping the early stages of galaxy
evolution.

This wealth of data has been contrasted against theoretical models of
galaxy formation \citep[see e.g.][]{LoFaro09, Lacey11, Cai14}. In
order to reproduce the observed LFs, all these studies favor a
scenario where considerable dust extinction needs to be considered, up
to the highest redshifts. \citet{Bouwens14}, however, argue that the
evolution of the UV continuum slope of high-z galaxies selected from
the CANDELS and HUDF fields requires that dust attenuation decreases
with decreasing luminosities and increasing redshift (being
considerably smaller at z$\sim$5-6 than at z$\sim$2-3). This implies
that the UV-derived SFR density accounts for most of the cosmic SFR at
z$>$4. Recent observations of 16 galaxies in the HUDF with the Atacama
Large Millimeter Array (ALMA) by \citet{Dunlop17} reinforce the
conclusion that at z$\gtrsim$4 most of the cosmic SFR is
unobscured. Recent results from \citet{RowanRobinson16} challenge,
however, this scenario. They use {\it Herschel} 500$\mu$m counts and
estimate SFR densities at 4$<$z$<$6 significantly higher than those
expected from UV. It is worth stressing that in this redshift range
SFR densities are still relatively uncertain, and are based on an
handful of exceptional objects with individual SFR estimates $>10^3
\msunyr$. Indeed, work by \citet{Bourne17} based on the {\it SCUBA-2}
Cosmology Legacy Survey, suggests a transition at z$\sim$4 from an
(almost) unobscured early phase of galaxy formation to later epochs
dominated by dust-obscured SFR, which the authors interpret as driven
by the formation of the most massive galaxies.

The redshift evolution of galaxies below the knee of the GSMF has long
been a problem for theoretical models of galaxy formation, which
typically predict these objects to form too early \citep{Fontanot09b,
  Weinmann12, Hirschmann12}. A number of recent studies
\citep{Henriques13, White15, Hirschmann16} point out that this problem
can be alleviated by modifications of the adopted stellar feedback
scheme. In detail, the most successful solutions invoke a combination
of a strong ejective feedback (in the form of strong stellar-driven
outflows) and a mass-dependent timescale for the re-accretion of the
ejected gas onto dark matter haloes. Most, if not all, of these
studies have focused on the $z<3$ GSMF, due to the more stringent
constraints available. \citet{LoFaro09} show that the problem of the
evolution of intermediate-to-low mass galaxies affects the predicted
shape of the high-z UV-LFs.

In this Letter, we investigate the impact of strong stellar-driven
outflows on galaxy properties beyond $z\sim4$, taking advantage of new
datasets and with the aim of presenting a coherent picture of galaxy
evolution over the widest redshift range available.

\section{Semi-analytic Model}\label{sec:models}
In this letter, we consider predictions from the model for GAlaxy
Evolution and Assembly ({\gaea} - \citealt{Hirschmann16}), which
represents an evolution of the \citet{DeLuciaBlaizot07} code. The new
model features significant improvements both in the treatment of
chemical enrichment (the code accounts for the non instantaneous
recycling of metals, gas and energy from asymptotic giant branch
stars, Type Ia and Type II Supernovae - \citealt{DeLucia14}) and in
the modeling of stellar feedback. In particular, \citet{Hirschmann16}
compared {\gaea} runs with different stellar feedback schemes to the
observed evolution of the GSMF. In this Letter, we consider two of
these schemes. The first one (``fiducial''), corresponds to the
standard ``energy-driven'' scheme implemented in \citet{DeLucia04b}
and \citet{DeLucia14}; the second one (H16F) corresponds to the
``FIRE'' stellar feedback implementation considered in
\citet{Hirschmann16}. In the latter model, gas reheating is
parametrized by using the fitting formulae discussed in
\citet{Muratov15}, based on the ``FIRE'' set of hydrodynamical
simulations \citep{Hopkins14}. The same physical dependencies are
assumed for the modeling of the rate of energy injection, while the
ejected gas mass (outside the dark matter haloes) is estimated
following energy conservation arguments as in \citet{Guo11}. Both the
ejection and reheating efficiencies are treated as free
parameters. Finally, we assume that the time-scale of gas
re-incorporation scales with the halo mass as assumed in
\citet{Henriques13}. \citet{Hirschmann16} show that an improved
modeling of both ejection and re-accretion is critical to reproduce
the evolution of galaxies below the knee of the GSMF in the redshift
range $0<z<3$, as well as the observed evolution of the galaxy mass-
and gas-metallicity relations up to $z\sim2$. The stellar feedback
strength in the H16F prescriptions is assumed to increase with
redshift so that outflows are stronger at higher redshifts. We stress
that in this work, we analyse model predictions on redshift and
stellar mass ranges that go well beyond both the original analysis by
\citet[z$\lesssim$4, $10^{10} \lesssim M_\star / \msun \lesssim
  10^{12}$]{Muratov15}, and the calibration in \citet[limited at
  z$\lesssim$3]{Hirschmann16}. We note that the H16F prescription is
not the only ejective feedback modeling able to reproduce the
evolution of the GSMF in {\gaea}, although being the only one able to
reproduce simultaneously the observed evolution of the
mass-metallicity relation. It is for this reason that we have elected
the H16F prescription as our reference model in the following work, as
well as in this study.

We couple {\gaea} with dark matter halo merger trees extracted from
the Millennium Simulation \citep[MS][]{Springel05}, a high resolution
cosmological simulation of a $\Lambda$CDM concordance model, with
parameters\footnote{Despite the values assumed for cosmological
  parameters are slightly different from the most recent determination
  \citep{Planck_cosmpar}, we do not expect this to affect
  significantly our conclusions \citep[see e.g.][]{Wang08}.}  assuming
a WMAP1 cosmology (i.e. $\Omega_\Lambda=0.75$, $\Omega_m=0.25$,
$\Omega_b=0.045$, $n=1$, $\sigma_8=0.9$, $H_0=73 \, {\rm
  km/s/Mpc}$). In order to extend model predictions to lower masses
and fainter luminosities we also consider runs based on the
Millennium-II Simulation \citep[MSII][]{BoylanKolchin09}, which
assumes the same cosmology, but a smaller cosmological volume and a
125 times better mass resolution. Throughout this Letter, we assume a
\citet{Chabrier03} initial mass function and we use the the stellar
population synthesis model from \citet{Bruzual03}.

\section{High-z LFs}\label{sec:results}
\begin{figure*}
  \plotone{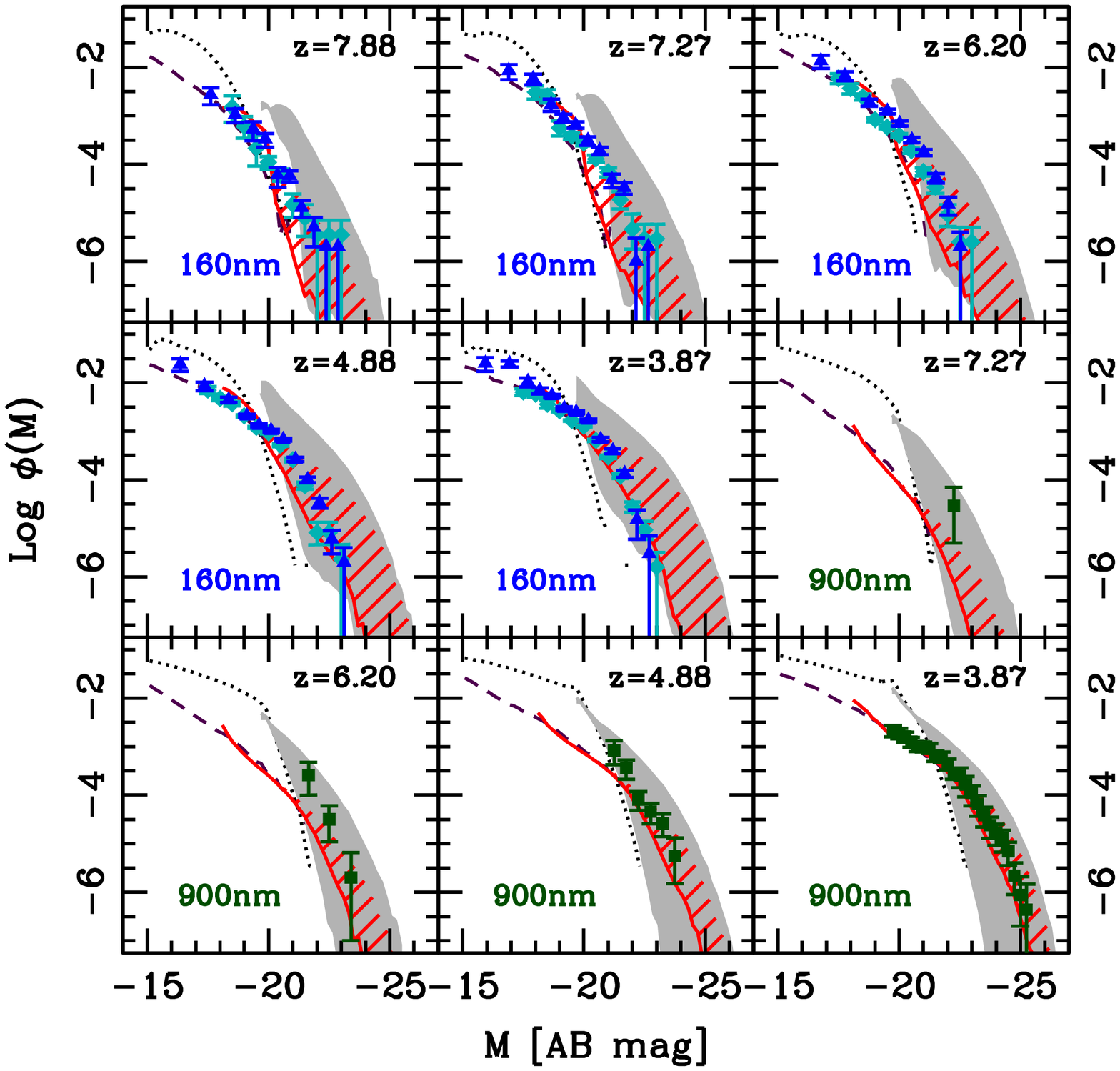}
  \caption{High-z (4$\lesssim$z$\lesssim$7) luminosity functions in
    different wavebands. Blue triangles and light blue diamonds
    correspond to the UV LFs from \citet{Bouwens15} and from
    \citet{Finkelstein15}, respectively. Green squares show the $900
          {\rm nm}$ LFs from \citet{Stefanon16}. The gray and the red
          hatched areas represent predictions from the Fiducial and
          H16F feedback implementations based on the MS. These areas
          have been determined from the LFs corresponding to the
          intrinsic and dust attenuated magnitudes. Dotted black and
          dashed dark red lines refer to predictions from the Fiducial
          and H16F feedback models (dust attenuated magnitudes) run on
          the MSII.}\label{fig:highzlf}
\end{figure*}
Fig.~\ref{fig:highzlf} presents the {\gaea} predicted LFs in the
rest-frame UV ($\sim 160 {\rm nm}$) and optical ($\sim 900 {\rm
  nm}$). Absolute UV magnitudes have been computed using a top-hat
filter centered at 160 $ {\rm nm}$ and 20 $ {\rm nm}$ wide; optical
magnitudes are computed in the $z_{ACS}$ filter. We compare our
predictions with a variety of high-z data coming from the recent
compilations of \citet{Bouwens15}, \citet{Finkelstein15} and
\citet{Stefanon16}. These data are all based on HST Legacy Fields, and
cover a wide wavelength range from the optical to the
Near-Infrared.

In order to estimate the effect of dust extinction in model galaxies,
we use the same approach as in \citet{DeLuciaBlaizot07}. Young stars
in dense birth clouds suffer larger extinction than evolved stars in
more diffuse cirrus. The age-dependent composite extinction curve is
scaled with the column density of dust in the disc, assuming the dust
mass is proportional to the metallicity. We further assume a ``slab''
geometry \citep{Devriendt99}, to provide an estimate of the total dust
attenuation in each source. This approach assumes a universal
composite extinction curve for model galaxies, as well as a
metallicity dependent normalization. Both are based on observations of
$z=0$ galaxies. Therefore, large uncertainties linger at the redshifts
considered in this study and might affect the comparison between
theoretical predictions and observational data. In order to keep them
under control, in Fig.~\ref{fig:highzlf} we consider both
dust-extincted and unextincted LFs at the relevant
wavelengths\footnote{The reference rest-frame wavelength varies
  slightly with redshift in data samples, due to the different filter
  sets used to select galaxies at different redshift.}: the gray
shaded (red hatched) area refers to predictions from the fiducial
(H16F) model, with the lower (upper) envelope corresponding to the
dust extincted (unextincted) LF. With respect to data, the unextincted
fiducial model tends to overestimate the LFs at all redshifts, and a
substantial amount of dust obscuration is needed to recover
observations. The extincted H16F feedback model reproduces the number
densities of UV and optical sources at z$\lesssim$5, but it
underpredicts the bright-end of the LFs at higher redshifts. On the
other hand, at $z>5$ the evolution of bright sources is better traced
by the intrinsic LFs. In order to recover the overall evolution of the
bright-end of the LF in the UV and optical bands, we thus have to
assume a decreasing importance of dust attenuation, in qualitative
agreement with \citet{Bouwens09}.

The magnitude range accessible with the MS is not wide enough to
sample the faint end of the LFs. In order to study the shape of the
LFs below the knee, we consider runs based on the MSII. We show
predictions for the dust attenuated LFs from the Fiducial and H16F
models (dotted black and dashed dark red lines, respectively) in
Fig.~\ref{fig:highzlf}. In both cases the convergence between the MSII
predictions and those based on the MS is satisfactory. Dust
attenuation is small for faint galaxies over the entire redshift
range, in agreement with the analysis of \citet{Bouwens09}. In
particular, the H16F run on the MSII reproduces the redshift evolution
of the faint-end of the UV-LFs up to $z$\lesssim$8$. In contrast, the
Fiducial model tends to over-predict the space density of faint UV
galaxies. {\it This result extends to the redshift range $4<z<7$ the
  evidence that strong stellar feedback represents a key ingredient to
  reproduce the observed evolution of the faint end of the LFs}. At
z$>$5 the available optical rest-frame data are not deep enough to
firmly discriminate between the two schemes. Over this redshift range,
the Fiducial model run on the MSII consistently predicts space
densities for faint sources larger than the H16F feedback scheme.

\section{Discussion}\label{sec:discussion}
\begin{figure*}
  \plotone{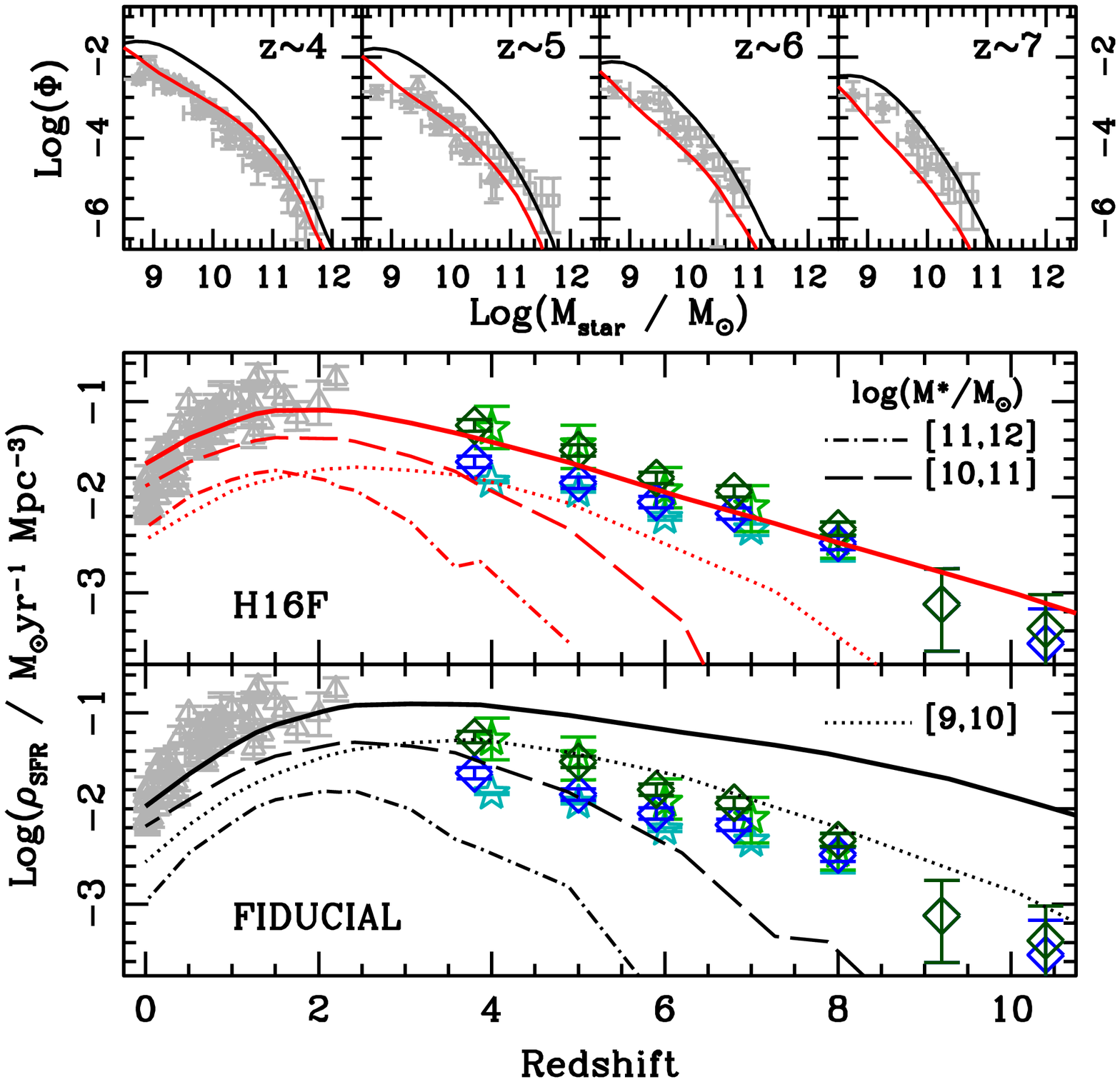}
  \caption{Upper panels: redshift evolution of the
    4$\lesssim$z$\lesssim$7 stellar mass function. Light gray points
    refer to data from \citet[asterisks]{Gonzalez11},
    \citet[squares]{Grazian15} and
    \citet[triangles]{Stefanon16}. Black and red solid lines represent
    predictions from the Fiducial and H16F feedback runs. Lower
    panels: cosmic star formation rate density. Blue and green dark
    diamonds reproduce SFR densities corrected and uncorrected for the
    effects of dust extinction as in \citet{Bouwens15}; light stars
    show data from \citet{Finkelstein15}. Light gray points correspond
    to the compilation of low-z determinations from
    \citet{Hopkins04}. Solid lines show the total SFR densities in the
    two models considered, while dotted, dashed and dot-dashed lines
    represent the contribution from galaxies in different stellar mass
    bins, as labeled.}\label{fig:phys}
\end{figure*}

We compare in Fig.~\ref{fig:phys} the observed evolution of the GSMF
and cosmic SFR density ($\rho_{\rm SFR}$), with model predictions. The
latter have been convolved with an estimate of the observational
errors, following \citet{Fontanot09b}, i.e. a log-normal error
distribution with amplitudes 0.25 and 0.3 for stellar masses and star
formation rates, respectively. In the upper panel, we compare the
predicted evolution of the GSMF at z$>$4 in the MS runs with
observational determination based on stellar masses derived from
spectral fitting techniques \citep{Gonzalez11, Grazian15} or from
mass-to-light ratios in the optical \citep{Stefanon16}. Results
confirm and extend to higher-redshifts the conclusions from
\citet{Hirschmann16}: models implementing the H16F feedback scheme are
able to reproduce the shape and redshift evolution of the GSMF, while
those based on the Fiducial scheme largely over-predict the number
densities of galaxies below the knee of the mass function. The H16F
feedback scheme is characterized by ejection rates that are larger
than those assumed in the fiducial model \citep[see e.g.
  Fig~4]{Hirschmann16}. This implies that large amounts of reheated
gas coupled with hot gas associated with dark matter haloes, are
ejected in a reservoir that is is assumed to be unavailable for
cooling. The observed evolution of the faint-end of the GSMF is then
recovered by also assuming a dependence of gas re-accretion time-scale
on halo mass. The agreement with data apparently worsens at the
highest redshifts considered, but we stress that at z$\gtrsim$6 model
predictions are computed on the closest snapshots available for the MS
and MSII (i.e. z=6.2 and z=7.2), that lie at a different redshift than
the mean redshift of observed samples (z$\sim$6 and z$\sim$7). This
mismatch could account for at least part of the disagreement.

In the bottom panel of Fig.~\ref{fig:phys}, we consider the $\rho_{\rm
  SFR}$ evolution. In order to perform a meaningful comparison with
high-z data from \citet{Bouwens15} or \citet{Finkelstein15} we only
consider the contribution of model galaxies with $M_{\rm 160nm}<-17$,
which roughly corresponds to $0.03 L^\star_{\rm z=3}$. Given this
faint integration limit, we consider in these panels predictions from
the MSII runs. Once again, the run implementing the H16F feedback
scheme provides an excellent agreement with observational
determinations up to z$\sim$10, while the Fiducial model over-predicts
the SFR density already at z$\sim$4. This confirms previous results
suggesting that strong stellar feedback regulates the early evolution
of $\rho_{\rm SFR}$ \citep[see e.g.][and references
  herein]{Vogelsberger13} We study the contribution of different
galaxy populations (binned in stellar mass) to the cosmic SFR. Our
results show that the main difference between the Fiducial and H16F
feedback schemes is seen in the evolution of the smallest
galaxies. The impact on $10^{11}<M_\star/\msun<10^{10}$ galaxies is
limited and mainly seen at z$\gtrsim$4; more massive galaxies are
those less affected by the different feedback schemes.

\section{Summary}\label{sec:summary}
In this letter we contrast predictions from our semi-analytic model
{\gaea} with the latest constraints on the evolution of the high-z
galaxies, using a combination of photometry (LFs in various bands) and
derived physical properties. In particular, we consider predictions
from our ``fiducial'' feedback scheme (that represents the standard
``energy-driven'' implementation from \citealt{DeLucia04b}) and from a
feedback scheme based on the results of hydrodynamical simulations
(H16F). Model predictions are compared to observational measurements
out to the highest redshifts probed by state-of-the-art surveys,
i.e. the edge of the epoch of reionization. Our results confirm and
extend the conclusions in \citet{Hirschmann16}, i.e. they clearly show
the need for strong stellar-driven outflows coupled with
mass-dependent re-accretion timescales in order to correctly reproduce
the evolution of the rest-frame UV and optical LFs over the redshift
range 4$<$z$<$7. In addition, {\gaea} runs implementing the H16F
feedback agree well with the evolution of the GSMF up to the highest
redshifts probed by the latest determinations (z$\sim$7) and with the
cosmic SFR derived by \citet{Bouwens15} \citep[or][]{Finkelstein15} up
to z$\sim$10.

We conclude that {\gaea} is able to reproduce the overall evolution of
the LFs, GSMF and cosmic SFR over the redshift range 0$<$z$<$10. It is
worth stressing that the good level of agreement shown in this letter
is obtained {\it without} any retuning of the feedback parameters
(which have been calibrated by \citealt{Hirschmann16} against lower
redshift observables). Our findings agree with recent results from the
MUFASA simulation suite \citet{Dave16}, that implements a kinetic
feedback scheme with scalings based on the parametrization provided in
\citet{Muratov15}. They show that the MUFASA runs are able to
reproduce the evolution of cosmic SFR and the GSMF at $z<4$ and the
evolution of its low-mass end slope up to z$\sim$6 (where the
simulated volume is too small to efficiently sample the high-mass end
of the GSMF).

A critical point of the analysis presented here is the treatment of
dust attenuation. In this letter, we consider both intrinsic and dust
attenuated magnitudes. Despite the simplified dust model adopted, we
show that the bright-ends of the corresponding LFs bracket the
observed LFs. Our results are in qualitative agreement with the
\citet{Bouwens09} inferences based on the UV continuum of the sources,
i.e. that dust attenuation should decrease at increasing redshifts,
becoming negligible at $z>5$. We note that a significant level of dust
attenuation is required by the Fiducial feedback scheme to match the
shape of the observed LFs, in agreement with previous studies
\citep{LoFaro09, Lacey11, Cai14}. Intrinsic and attenuated LFs
converge below the knee of the LFs, again in qualitative agreement
with the estimates by \citet{Bouwens09} of a lower dust attenuation
for fainter sources.

The predicted level of dust attenuation at high redshift thus
represents a powerful discriminant between different feedback
schemes. Therefore, further insight in our understanding of galaxy
evolution at z$\gtrsim$4 is tightly connected to a better description
of this key aspect. Form a theoretical perspective, the implementation
of a self-consistent dust treatment in theoretical models of galaxy
evolution, able to follow dust production and destruction alongside
with galaxy assembly (see e.g., the recent work by
\citealt{Popping16}) represents a promising avenue. In addition, dust
provides an important channel for molecular hydrogen formation, and
can therefore play a crucial role in regulating star formation at
different cosmic epochs. This modeling is beyond the aims of this
letter, and will be the subject of future work. Here, we just
highlight that our conclusions will change dramatically if a
substantial fraction of the SFR density is related to highly obscured
objects, as some studies suggest at $3<z<6$
\citep{RowanRobinson16}. In order to clearly assess the contribution
of dusty sources to the SFR density at $z>4$, and to distinguish
between different incarnations for stellar feedback, forthcoming or
proposed facilities (like the {\it James Webb Space Telescope} - JWST
- or the {\it SPace Infrared telescope for Cosmology and Astrophysics}
- SPICA) will be of paramount importance.


\section*{Acknowledgments}

Galaxy catalogs from our new {\gaea} model (implementing the H16F
feedback scheme) based on the Millennium merger trees will be publicly
available at {\it http://www.mpa-garching.mpg.de/millennium}. FF and
GDL acknowledge financial support from the MERAC foundation and from
the PRIN INAF 2014 ``Glittering kaleidoscopes in the sky: the
multifaceted nature and role of Galaxy Clusters.'' MH acknowledges
financial support from the European Research Council via an Advanced
Grant under grant agreement no. 321323 (NEOGAL).

\bibliographystyle{aasjournal}
\bibliography{fontanot}

\begin{thebibliography}{}
\expandafter\ifx\csname natexlab\endcsname\relax\def\natexlab#1{#1}\fi
\providecommand{\url}[1]{\href{#1}{#1}}

\bibitem[{{Bourne} {et~al.}(2017){Bourne}, {Dunlop}, {Merlin}, {Parsa},
  {Schreiber}, {Castellano}, {Conselice}, \& {Coppin}}]{Bourne17}
{Bourne}, N., {Dunlop}, J.~S., {Merlin}, E., {et~al.} 2017, \mnras,
  arXiv:1607.04283

\bibitem[{{Bouwens} {et~al.}(2009){Bouwens}, {Illingworth}, {Franx}, {Chary},
  {Meurer}, {Conselice}, {Ford}, {Giavalisco}, \& {van Dokkum}}]{Bouwens09}
{Bouwens}, R.~J., {Illingworth}, G.~D., {Franx}, M., {et~al.} 2009, \apj, 705,
  936

\bibitem[{{Bouwens} {et~al.}(2014){Bouwens}, {Illingworth}, {Oesch},
  {Labb{\'e}}, {van Dokkum}, {Trenti}, {Franx}, {Smit}, {Gonzalez}, \&
  {Magee}}]{Bouwens14}
{Bouwens}, R.~J., {Illingworth}, G.~D., {Oesch}, P.~A., {et~al.} 2014, \apj,
  793, 115

\bibitem[{{Bouwens} {et~al.}(2015){Bouwens}, {Illingworth}, {Oesch}, {Trenti},
  {Labb{\'e}}, {Bradley}, {Carollo}, {van Dokkum}, {Gonzalez}, {Holwerda},
  {Franx}, {Spitler}, {Smit}, \& {Magee}}]{Bouwens15}
---. 2015, \apj, 803, 34

\bibitem[{{Bouwens} {et~al.}(2016){Bouwens}, {Oesch}, {Labb{\'e}},
  {Illingworth}, {Fazio}, {Coe}, {Holwerda}, {Smit}, {Stefanon}, {van Dokkum},
  {Trenti}, {Ashby}, {Huang}, {Spitler}, {Straatman}, {Bradley}, \&
  {Magee}}]{Bouwens16}
{Bouwens}, R.~J., {Oesch}, P.~A., {Labb{\'e}}, I., {et~al.} 2016, \apj, 830, 67

\bibitem[{{Boylan-Kolchin} {et~al.}(2009){Boylan-Kolchin}, {Springel}, {White},
  {Jenkins}, \& {Lemson}}]{BoylanKolchin09}
{Boylan-Kolchin}, M., {Springel}, V., {White}, S.~D.~M., {Jenkins}, A., \&
  {Lemson}, G. 2009, \mnras, 398, 1150

\bibitem[{{Bruzual} \& {Charlot}(2003)}]{Bruzual03}
{Bruzual}, G., \& {Charlot}, S. 2003, \mnras, 344, 1000

\bibitem[{{Cai} {et~al.}(2014){Cai}, {Lapi}, {Bressan}, {De Zotti}, {Negrello},
  \& {Danese}}]{Cai14}
{Cai}, Z.-Y., {Lapi}, A., {Bressan}, A., {et~al.} 2014, \apj, 785, 65

\bibitem[{{Chabrier}(2003)}]{Chabrier03}
{Chabrier}, G. 2003, \apjl, 586, L133

\bibitem[{{Dav{\'e}} {et~al.}(2016){Dav{\'e}}, {Thompson}, \&
  {Hopkins}}]{Dave16}
{Dav{\'e}}, R., {Thompson}, R., \& {Hopkins}, P.~F. 2016, \mnras, 462, 3265

\bibitem[{{De Lucia} \& {Blaizot}(2007)}]{DeLuciaBlaizot07}
{De Lucia}, G., \& {Blaizot}, J. 2007, \mnras, 375, 2

\bibitem[{{De Lucia} {et~al.}(2004){De Lucia}, {Kauffmann}, \&
  {White}}]{DeLucia04b}
{De Lucia}, G., {Kauffmann}, G., \& {White}, S.~D.~M. 2004, \mnras, 349, 1101

\bibitem[{{De Lucia} {et~al.}(2014){De Lucia}, {Tornatore}, {Frenk}, {Helmi},
  {Navarro}, \& {White}}]{DeLucia14}
{De Lucia}, G., {Tornatore}, L., {Frenk}, C.~S., {et~al.} 2014, \mnras, 445,
  970

\bibitem[{{Devriendt} {et~al.}(1999){Devriendt}, {Guiderdoni}, \&
  {Sadat}}]{Devriendt99}
{Devriendt}, J.~E.~G., {Guiderdoni}, B., \& {Sadat}, R. 1999, \aap, 350, 381

\bibitem[{{Dunlop} {et~al.}(2017){Dunlop}, {McLure}, {Biggs}, {Geach},
  {Micha{\l}owski}, {Ivison}, {Rujopakarn}, \& {van Kampen}}]{Dunlop17}
{Dunlop}, J.~S., {McLure}, R.~J., {Biggs}, A.~D., {et~al.} 2017, \mnras, 466,
  861

\bibitem[{{Finkelstein} {et~al.}(2015){Finkelstein}, {Ryan}, {Papovich},
  {Dickinson}, {Song}, {Somerville}, {Ferguson}, {Salmon}, \&
  et~al.}]{Finkelstein15}
{Finkelstein}, S.~L., {Ryan}, Jr., R.~E., {Papovich}, C., {et~al.} 2015, \apj,
  810, 71

\bibitem[{{Fontanot} {et~al.}(2014){Fontanot}, {Cristiani}, {Pfrommer},
  {Cupani}, \& {Vanzella}}]{Fontanot14a}
{Fontanot}, F., {Cristiani}, S., {Pfrommer}, C., {Cupani}, G., \& {Vanzella},
  E. 2014, \mnras, 438, 2097

\bibitem[{{Fontanot} {et~al.}(2009){Fontanot}, {De Lucia}, {Monaco},
  {Somerville}, \& {Santini}}]{Fontanot09b}
{Fontanot}, F., {De Lucia}, G., {Monaco}, P., {Somerville}, R.~S., \&
  {Santini}, P. 2009, \mnras, 397, 1776

\bibitem[{{Gonz{\'a}lez} {et~al.}(2011){Gonz{\'a}lez}, {Labb{\'e}}, {Bouwens},
  {Illingworth}, {Franx}, \& {Kriek}}]{Gonzalez11}
{Gonz{\'a}lez}, V., {Labb{\'e}}, I., {Bouwens}, R.~J., {et~al.} 2011, \apjl,
  735, L34

\bibitem[{{Grazian} {et~al.}(2015){Grazian}, {Fontana}, {Santini}, {Dunlop},
  {Ferguson}, {Castellano}, {Amorin}, \& {Ashby}}]{Grazian15}
{Grazian}, A., {Fontana}, A., {Santini}, P., {et~al.} 2015, \aap, 575, A96

\bibitem[{{Guo} {et~al.}(2011){Guo}, {White}, {Boylan-Kolchin}, {De Lucia},
  {Kauffmann}, {Lemson}, {Li}, {Springel}, \& {Weinmann}}]{Guo11}
{Guo}, Q., {White}, S., {Boylan-Kolchin}, M., {et~al.} 2011, \mnras, 413, 101

\bibitem[{{Henriques} {et~al.}(2013){Henriques}, {White}, {Thomas}, {Angulo},
  {Guo}, {Lemson}, \& {Springel}}]{Henriques13}
{Henriques}, B.~M.~B., {White}, S.~D.~M., {Thomas}, P.~A., {et~al.} 2013,
  \mnras, 431, 3373

\bibitem[{{Hirschmann} {et~al.}(2016){Hirschmann}, {De Lucia}, \&
  {Fontanot}}]{Hirschmann16}
{Hirschmann}, M., {De Lucia}, G., \& {Fontanot}, F. 2016, \mnras, 461, 1760

\bibitem[{{Hirschmann} {et~al.}(2012){Hirschmann}, {Somerville}, {Naab}, \&
  {Burkert}}]{Hirschmann12}
{Hirschmann}, M., {Somerville}, R.~S., {Naab}, T., \& {Burkert}, A. 2012,
  \mnras, 426, 237

\bibitem[{{Hopkins}(2004)}]{Hopkins04}
{Hopkins}, A.~M. 2004, \apj, 615, 209

\bibitem[{{Hopkins} {et~al.}(2014){Hopkins}, {Kere{\v s}}, {O{\~n}orbe},
  {Faucher-Gigu{\`e}re}, {Quataert}, {Murray}, \& {Bullock}}]{Hopkins14}
{Hopkins}, P.~F., {Kere{\v s}}, D., {O{\~n}orbe}, J., {et~al.} 2014, \mnras,
  445, 581

\bibitem[{{Lacey} {et~al.}(2011){Lacey}, {Baugh}, {Frenk}, \&
  {Benson}}]{Lacey11}
{Lacey}, C.~G., {Baugh}, C.~M., {Frenk}, C.~S., \& {Benson}, A.~J. 2011,
  \mnras, 412, 1828

\bibitem[{{Lo Faro} {et~al.}(2009){Lo Faro}, {Monaco}, {Vanzella}, {Fontanot},
  {Silva}, \& {Cristiani}}]{LoFaro09}
{Lo Faro}, B., {Monaco}, P., {Vanzella}, E., {et~al.} 2009, \mnras, 399, 827

\bibitem[{{Muratov} {et~al.}(2015){Muratov}, {Kere{\v s}},
  {Faucher-Gigu{\`e}re}, {Hopkins}, {Quataert}, \& {Murray}}]{Muratov15}
{Muratov}, A.~L., {Kere{\v s}}, D., {Faucher-Gigu{\`e}re}, C.-A., {et~al.}
  2015, \mnras, 454, 2691

\bibitem[{{Planck Collaboration XVI}(2014)}]{Planck_cosmpar}
{Planck Collaboration XVI}. 2014, \aap, 571, A16

\bibitem[{{Popping} {et~al.}(2016){Popping}, {Somerville}, \&
  {Galametz}}]{Popping16}
{Popping}, G., {Somerville}, R.~S., \& {Galametz}, M. 2016, ArXiv e-prints
  (arXiv:1609.08622), arXiv:1609.08622

\bibitem[{{Robertson} {et~al.}(2013){Robertson}, {Furlanetto}, {Schneider},
  {Charlot}, {Ellis}, {Stark}, {McLure}, {Dunlop}, {Koekemoer}, {Schenker},
  {Ouchi}, {Ono}, {Curtis-Lake}, {Rogers}, {Bowler}, \&
  {Cirasuolo}}]{Robertson13}
{Robertson}, B.~E., {Furlanetto}, S.~R., {Schneider}, E., {et~al.} 2013, \apj,
  768, 71

\bibitem[{{Rowan-Robinson} {et~al.}(2016){Rowan-Robinson}, {Oliver}, {Wang},
  {Farrah}, {Clements}, {Gruppioni}, {Marchetti}, {Rigopoulou}, \&
  {Vaccari}}]{RowanRobinson16}
{Rowan-Robinson}, M., {Oliver}, S., {Wang}, L., {et~al.} 2016, \mnras, 461,
  1100

\bibitem[{{Springel} {et~al.}(2005){Springel}, {White}, {Jenkins}, {Frenk},
  {Yoshida}, {Gao}, {Navarro}, {Thacker}, {Croton}, {Helly}, {Peacock}, {Cole},
  {Thomas}, {Couchman}, {Evrard}, {Colberg}, \& {Pearce}}]{Springel05}
{Springel}, V., {White}, S.~D.~M., {Jenkins}, A., {et~al.} 2005, \nat, 435, 629

\bibitem[{{Stefanon} {et~al.}(2016){Stefanon}, {Bouwens}, {Labb{\'e}},
  {Muzzin}, {Marchesini}, {Oesch}, \& {Gonzalez}}]{Stefanon16}
{Stefanon}, M., {Bouwens}, R.~J., {Labb{\'e}}, I., {et~al.} 2016, ArXiv
  e-prints (arXiv:1611.09354), arXiv:1611.09354

\bibitem[{{Steidel} {et~al.}(1996){Steidel}, {Giavalisco}, {Pettini},
  {Dickinson}, \& {Adelberger}}]{Steidel96}
{Steidel}, C.~C., {Giavalisco}, M., {Pettini}, M., {Dickinson}, M., \&
  {Adelberger}, K.~L. 1996, \apjl, 462, L17

\bibitem[{{Vogelsberger} {et~al.}(2013){Vogelsberger}, {Genel}, {Sijacki},
  {Torrey}, {Springel}, \& {Hernquist}}]{Vogelsberger13}
{Vogelsberger}, M., {Genel}, S., {Sijacki}, D., {et~al.} 2013, \mnras, 436,
  3031

\bibitem[{{Wang} {et~al.}(2008){Wang}, {De Lucia}, {Kitzbichler}, \&
  {White}}]{Wang08}
{Wang}, J., {De Lucia}, G., {Kitzbichler}, M.~G., \& {White}, S.~D.~M. 2008,
  \mnras, 384, 1301

\bibitem[{{Weinmann} {et~al.}(2012){Weinmann}, {Pasquali}, {Oppenheimer},
  {Finlator}, {Mendel}, {Crain}, \& {Macci{\`o}}}]{Weinmann12}
{Weinmann}, S.~M., {Pasquali}, A., {Oppenheimer}, B.~D., {et~al.} 2012, \mnras,
  426, 2797

\bibitem[{{White} {et~al.}(2015){White}, {Somerville}, \& {Ferguson}}]{White15}
{White}, C.~E., {Somerville}, R.~S., \& {Ferguson}, H.~C. 2015, \apj, 799, 201

\end{thebibliography}

\end{document}